\documentclass[aps,preprint,fleqn]{revtex4}

\begin{document}

\title{Proton Zemach radius from measurements of the hyperfine splitting of
hydrogen and muonic hydrogen}

\author{Arnaud Dupays}
\author{Alberto Beswick}
\author{Bruno Lepetit}
\author{Carlo Rizzo}
\affiliation{LCAR--IRSAMC, Universit\'{e} Paul Sabatier,
118 Route de Narbonne, 31062 Toulouse cedex 04, France}

\author{Dimitar Bakalov}
\affiliation{Institute for Nuclear Research and Nuclear Energy,
Tsarigradsko chauss\'{e}e 72, Sofia 1784, Bulgaria}

\begin{abstract}
While measurements of the hyperfine structure of hydrogen-like
atoms are traditionally regarded as test of bound-state QED, we
assume that theoretical QED predictions are accurate and discuss
the information about the electromagnetic structure of protons
that could be extracted from the experimental values of the ground
state hyperfine splitting in hydrogen and muonic hydrogen.
Using recent theoretical results on the proton polarizability effects
and the experimental
hydrogen hyperfine splitting
we obtain for the Zemach radius of the proton the value
1.040(16) fm. We compare it to the various theoretical estimates
the uncertainty of which is shown to be larger that 0.016 fm.
This
point of view gives quite convincing arguments in support of
projects to measure the hyperfine splitting of muonic hydrogen.
\end{abstract}
\maketitle

\section{Introduction}

The hyperfine splitting of the ground state of the hydrogen atom
is among the most accurately measured quantities \cite{H-hfs1,
H-hfs2}:
\begin{equation}
\Delta E^{\rm HFS}_{exp}=1\,420\,405\,751.\,7667\pm0.0009 {\rm
Hz}, \label{essen}
\end{equation}
The relative experimental uncertainty in Eq.~\ref{essen} does not exceed
$10^{-12}$. The theoretical predictions for $\Delta E^{\rm HFS}$,
based on QED, are less accurate. This is partly due to the
computational difficulties which increase very fast for the higher
order terms in the perturbative expansion in powers of $\alpha$ and $(Z\alpha)$ \cite{sapir,eides},
and to the limited precision of the fundamental constants involved
(the Rydberg known to $10^{-11}$, electron to proton mass ratio and
$\alpha$ known to parts of $10^{-8}$ \cite{eides}).
The main uncertainties come,
however, from the insufficient knowledge of the structure of
protons. Because of this, the comparison of theoretical results with
the experimental value of Eq.~\ref{essen} can not test QED beyond
the contribution of proton polarizability effects of the order of a few
ppm.
To perform a more precise test of QED, therefore, either additional
information on the electromagnetic structure of protons should be
used, or the comparison should be done between theoretical and experimental
results on the hyperfine splitting in hydrogen-like bound states of charged
point-like leptons.
Muonium is most appropriate for the latter; indeed, the
recent measurements of the hyperfine
splitting of the
ground state of muonium \cite{muonium-hfs-exp} have been shown to agree
with theory up to $0.5\,10^{-7}$ (see \cite{eides} and references therein),
so that the correctness of QED results about the hyperfine splitting of
hydrogen-like atoms has been experimentally confirmed at least with the
same precision.
As for the use of information
on the electromagnetic structure of
protons from other sources in attempts to reduce
the theoretical uncertainty of $\Delta E^{\rm HFS}$,
until recently there was no theoretical progress in this direction,
and the only realistic idea for years seemed to be to
estimate the proton structure contributions
from complementary measurements of the hyperfine splitting in
muonic hydrogen \cite{pla,wein}.
We are now going to critically analyze this idea in the context of
recent theoretical results on the polarizability of protons
\cite{faustov-HHFS3,cherednikova}, of the
development of new precision spectroscopy instrumentation
\cite{kottmann-2000}, and of our
better understanding of the dominating proton structure contributions
to the hyperfine splitting in hydrogen-like atoms.
This will lead us to the alternative point of view to look at the
hyperfine splitting measurements in hydrogen and muonic hydrogen
as measurements of the Zemach
radius of the proton by assuming
that all QED predictions are credible.
This way we shall obtain a ``first experimental'' value of
the proton Zemach radius from the hyperfine splitting of
hydrogen which may be compared to theoretical values based on
different proton form factor fits.

\section{Hyperfine splitting of hydrogen atom ground state}

To analyze the various sources of uncertainty in the
theoretical value of $\Delta E^{\rm HFS}$ we
put it in the traditional form
\cite{bodwin}:
\begin{equation}
\Delta E^{\rm HFS}_{th}=E^{\rm F}\left(1+\delta^{\rm QED}
+\delta^{\rm str}\right) \label{qed+str}
\end{equation}
where $E^{\rm F}$ is the Fermi splitting \cite{fermi} expressed in
terms of the electron and proton masses $m_e,m_p$ and the dipole
magnetic moment of the proton $\mu_p$:
\begin{equation}
E^{\rm F}=\frac{8}{3}\alpha^4 c^2 \frac{m_e^2
m_p^2}{(m_e+m_p)^3}\mu_p, \label{E^F}
\end{equation}
while $\delta^{\rm QED}$ and $\delta^{\rm str}$ are correction
terms related to higher order QED effects and to proton electromagnetic
structure due to strong interactions.
(Eq.~\ref{qed+str} is only correct
in the leading order, since higher orders QED and structure
effects mix up \cite{karshen-mix}.)
Up to terms of order $O(\alpha^3)$ (without
distinguishing $\alpha$ from $(Z\alpha)$),  $\delta^{\rm QED}$ is
given by \cite{sapir,eides}:
\begin{equation}
\delta^{\rm QED}=a_e+\frac{3}{2}\alpha^2+\alpha^2(\log
2-\frac{5}{2})-\frac{8\alpha^3}{3\pi}\log \alpha(\log \alpha-\log
4+\frac{281}{480})+18.984\times\frac{\alpha^3}{\pi}+\ldots
\label{qed}
\end{equation}
where $a_e$ is the anomalous magnetic moment of the electron. Note
that the expression for $\delta^{\rm QED}$ does not involve the
mass ratio $m_e/m_p$;
all terms which depend on proton mass
or come from strong interactions are included in $\delta^{\rm str}$.
In turn, $\delta^{\rm str}$ splits into a ``static''
part $\delta^{\rm rigid}$ that accounts for the  elastic
electromagnetic form factors of the proton and can be calculated
using data from elastic scattering experiments, a part
$\delta^{\rm pol}$ that comes from the internal dynamics of the
proton and could only be evaluated using data on inelastic
processes with protons, and a part $\delta^{\rm hvp}$ describing
the strong interaction effects outside the proton, such as
hadron vacuum polarization:
$\delta^{\rm str}= \delta^{\rm rigid} +\delta^{\rm pol}
+\delta^{\rm hvp}$.
Two types of ``static'' proton structure corrections are
incorporated in $\delta^{\rm rigid}$, associated with the spatial
distribution of the charge and magnetic moment within the proton
and with recoil effects, respectively: $\delta^{\rm
rigid}=\delta^{\rm Zemach}+\delta^{\rm recoil}$.
The former has been calculated by Zemach \cite{zemach} and may be
put in the form \cite{cherednikova}:
\begin{equation}
\delta^{\rm Zemach}=\frac{2\alpha
m_{ep}}{\pi^2}\int\frac{d^3p}{(\mbox{\bf p}^2+\alpha^2 m_{ep}^2)^2}
\left(\frac{1}{\mu_p}G_E(-\mbox{\bf p}^2)G_M(-\mbox{\bf p}^2)-1\right)
=-2\alpha m_{ep} R_p+O(\alpha^3) \label{zemach}
\end{equation}
where $m_{ep}=m_e m_p/(m_e+m_p)$ and $R_p$ is the first moment of
the convolution of the proton charge and magnetic moment
distributions, also known as Zemach radius of the proton. The
recoil correction $\delta^{\rm recoil}$ denotes the contribution
of all terms which depend on the ratio $m_e/m_p$; for sake of
simplicity we skip here the
rather lengthy explicit
expression
of $\delta^{\rm recoil}$ which may be
found in \cite{bodwin}.
The proton polarizability correction $\delta^{\rm pol}$
and the hadron vacuum polarization correction have been
evaluated recently \cite{faustov-HHFS3,faustov-HHFS2}, using the
available data on the proton polarized structure functions and
on electron-positron annihilation into hadrons.

Eq.~\ref{qed+str} may now be put in a more detailed form:
\begin{equation}
\Delta E^{\rm HFS}_{th}=E^{\rm F}\left(1+\delta^{\rm QED}
+\delta^{\rm Zemach}+\delta^{\rm recoil}+\delta^{\rm pol}+
\delta^{\rm hvp}
\right) \label{qed+str-det}
\end{equation}
It is important for what follows to review the order of
magnitude of the various terms in the right-hand side of Eq.~\ref{qed+str-det}
and briefly discuss the uncertainty of each of them.
The uncertainty of $E^{\rm F}$ is due to the uncertainty of $m_e$,
$m_p$, $\mu_p$ and $\alpha$ and does not exceed 0.01 ppm
\cite{partprop}. $\delta^{\rm QED}$ is dominated by the anomalous
magnetic moment of the electron; the uncertainty of the term comes
from the uncalculated terms of relative order $O(\alpha^4)$ and higher and
is estimated not to exceed 0.001 ppm \cite{eides}.
$\delta^{\rm rigid}$ is close to 40 ppm
(\cite{bodwin} and references therein); its uncertainty -- 2-3
ppm -- is due to the uncertainty of the actual values of the
proton radii and is of the order of 5\%.
The recoil term $\delta^{\rm recoil}\approx 5.68$ ppm \cite{sapir}
adds little uncertainty to $\delta^{\rm rigid}$.
As for the proton polarizability correction $\delta^{\rm pol}$,
until recently there existed only the upper limit $\delta^{\rm
pol}<4$ ppm \cite{sapir}, obtained
by ascribing the whole discrepancy between $\Delta
E^{\rm HFS}_{exp}$ and $\Delta E^{\rm HFS}_{th}$ to the
contribution from $\delta^{\rm pol}$.
The present value of the proton
polarizability correction, calculated on the ground of
experimental data on the polarized structure function
\cite{faustov-HHFS3}, is 1.6 $\pm$ 0.6 ppm.
The hadron vacuum polarization correction
$\delta^{\rm hvp}\sim10^{-8}$ \cite{faustov-HVPmu} is much
too small.
The overall uncertainty of $\Delta E^{\rm HFS}_{th}$ is therefore
of the order of 2-3 ppm and is entirely due to proton structure
effects.
All discussed quantities are summarized in the leftmost two
columns of Table \ref{tab1}; the righmost two columns of the Table contain
the numerical values of the corresponding quantities for muonic hydrogen and will be discussed in next section.

\begin{table}[h]
\label{tab1}
\begin{center}
\caption{Magnitude and uncertainty of the contributions to the
hyperfine splitting of the ground state of hydrogen and muonic hydrogen
from various correction terms.}
\begin{tabular}{|l|rr|rr|}
\cline{2-5}
 \multicolumn{1}{c}{}& \multicolumn{2}{|c|}{Hydrogen} &
\multicolumn{2}{c|}{Muonic Hydrogen} \\ \cline{2-5}
 \multicolumn{1}{c|}{} & magnitude & uncertainty & magnitude & uncertainty\\ \hline
$E^{\rm F}$ & 1420 MHz & 0.01 ppm
 & 182.443 meV & 0.1 ppm \\
 \hline\hline
$\delta^{\rm QED}$ & $1.16\,10^{-3}$  & $<0.001\,10^{-6}$ & $1.16\,10^{-3}$ &
$10^{-6}$\\
%
%
$\delta^{\rm rigid}$ & $39\,10^{-6}$ & $2\,10^{-6}$ & $7.5\,10^{-3}$ & $0.1\,10^{-3}$ \\
$\delta^{\rm recoil}$ & $6\,10^{-6}$ & $10^{-8}$ & $1,7\,10^{-3}$ & $10^{-6}$\\
$\delta^{\rm pol}$ &  $1.4\,10^{-6}$ & $0.6\,10^{-6}$ & $0.46\,10^{-3}$ & $0.08\,10^{-3}$\\
$\delta^{\rm hvp}$ & $10^{-8}$ & $10^{-9}$ & $0.02\,10^{-3}$ &
$0.002\,10^{-3}$ \\
\hline
\end{tabular}
\end{center}
\end{table}

As already pointed out, the
challenging comparison of $\Delta E^{\rm HFS}_{th}$ and $\Delta
E^{\rm HFS}_{exp}$ for hydrogen can not be regarded as a test of the QED
contributions $\delta^{\rm QED}$ of order $O(\alpha^3)$ and higher
because of the significant overall theoretical uncertainty.
We adopt instead an
alternative point of view to assume that
the theoretical values of $\delta^{\rm QED}$,
$\delta^{\rm recoil}$, $\delta^{\rm hvp}$ and
$\delta^{\rm pol}$ are
accurate and use the experimental data to determine
the Zemach radius of the
proton $R_p$ as:
\begin{equation}
R_p=-\left(
\Delta E^{\rm HFS}_{exp}/E^{\rm F}-1-\delta^{\rm QED}
-\delta^{\rm recoil}-\delta^{\rm pol}-
\delta^{\rm hvp}
\right)/(2m_{ep}\alpha)
\label{zemexp}
\end{equation}
The above assumption is justifiable since all four correction terms
are objects of QED, the only difference of $\delta^{\rm hvp}$ and
$\delta^{\rm pol}$ from the former two being that their evaluation
requires the use of additional phenomenological information
beyond first principles.
From Eq.~\ref{zemexp} we get the experimental value
$R_p=1.040(16)$ fm, where the
uncertainty $\pm0.016$ fm comes from the theoretical uncertainty
of $\delta^{\rm pol}$ \cite{faustov-HHFS3}.

The Zemach radius of the proton is defined in terms of an integral
of the charge and magnetic form factors of the proton $G_E(k)$
and $G_E(k)$ over space-like transfer  momenta
$k, k^2=-{\bf k}^2$ (see Eq.~\ref{zemach}, or equivalently,
by the first moment of the convolution of the charge and magnetic moment
distributions $\rho_E(r)$ and $\rho_M(r)$ in coordinate space \cite{zemach}:
\begin{equation}
R_p=\int d^3r\,r\int d^3r'\, \rho_E({\bf r}-{\bf r}')\, \rho_M({\bf r}')
\end{equation}
The directly observable quantity which is most
sensitive to the Zemach radius of the proton is the hyperfine splitting
of bound systems involving protons (compare to the Lamb shift
which is related to the proton r.m.s. charge radius
\cite{pachucki-FSZmH}.)
The experimental value of $R_p$
sets important restrictions on the theoretical models of proton
electromagnetic structure and, in particular, on the parametrization
of proton form factors, in terms of which the theoretical values are
calculated.
Consider as an example the values of $R_p$ calculated from
Eq.~\ref{zemach} using a few popular approximations of the
proton form factors.
Numerical calculations give
$R_p$=1.02 fm for the dipole fit, and
$R_p$=1.067 fm for the fit of \cite{simon}; unfortunately,
no information on the uncertainty of the parameters of the fit
is available, and no conclusions  could be made on the compatibility of
these values with the ''experimental'' one.
Both fits are consistent with $\mu_p G_E(-{\bf k}^2)/G_M(-{\bf k}^2)
\approx1$ for ${\bf k}^2<5$ GeV$^2$/c$^2$.
To account for the recent experimental results on the form factor ratio
\cite{gayou},
we also evaluated $R_p$ by using the Simon's fit for either the charge
or magnetic form factor and expressing the other one using
the relation $\mu_p G_E(-{\bf k}^2)/G_M(-{\bf k}^2)=
1-0.13\,({\bf k}^2-0.04)$ \cite{gayou}, and got
$R_p$=1.060 fm and $R_p$=1.073 fm respectively.
Though preliminary, these estimates
show that the current theoretical uncertainty of $R_p$
significantly exceeds the experimental one, and that
the experimental results on the proton
Zemach radius may be used as a test for the quality of models of the
proton in the limit of low transfer momenta.

\section{Hyperfine splitting of muonic hydrogen atom ground
state}

Muonic hydrogen is the only other hydrogen-like atom in which the
hyperfine splitting of the ground state could be measured with
high precision.
Due to the large muon mass $m_{\mu}/m_e\approx 2\,10^{2}$, the binding
energy of the ground state of muonic hydrogen is of the order of
200 Ry, and the radius of the muon orbit is $\sim a_0/200$ so that
the energy levels of muonic hydrogen
are more ``sensitive'' to the details of the
proton structure than the levels of normal hydrogen.
The expressions in Eqs.~\ref{qed+str},\ref{E^F} and \ref{zemach}
for the hyperfine splitting of a hydrogen-like atom and the
various contributions to it apply for muonic hydrogen as well;
however, the energy scale and the relative size of the various
terms differ significantly from the hydrogen case (see the
rightmost two columns of Table \ref{tab1}.) The Fermi splitting
now is 183 meV that corresponds to a hyperfine transition
wavelength of 6.1 $\mu$m. The explicit form of the higher order terms
in $\delta^{\rm QED}$ may differ from Eq.~\ref{qed} since
different momenta are expected to give the main contribution in
loop integrations, while proton structure modifies the proton
voerteces. We do not know of any published explicit
expression of $\delta^{\rm QED}$ for the ground state of muonic
hydrogen. To our opinion the lack of interest in the topic is due
to the unclear perspectives of an experimental verification of the
theoretical results. Indeed, the analogous contributions to the
Lamb shift of muonic hydrogen were evaluated with a very high
accuracy \cite{pachucki-LSmH,kinoshita1} as soon as the proposal for
the  experiment \cite{kottmann}  was about to be put forward. We
are therefore convinced that the uncertainty of $\delta^{\rm QED}$
may be brought down to 0.1 ppm if necessary. The evaluation of the
recoil terms $\delta^{\rm recoil}$ may not be that easy since the
mass ratio $m_{\mu}/m_p\sim0.11$ is much larger compared to
hydrogen, and terms of order $O(\alpha(m_{\mu}/m_p)^n),\ n=2,3,4$,
are all expected to contribute by more than 1 ppm. Most
appropriate in this case might be an essentially two-body approach
based on the quasipotential equation \cite{todorov,faustov-QP};
we assume that the theoretical uncertainty may be brought
down below $10^{-6}$ this way, and put 1 ppm in the rightmost column
of Table \ref{tab1}. Since
we do not know of any published result on $\delta^{\rm recoil}$,
we take as estimate of the magnitude of the recoil effects the
formula for the leading recoil term in muonium of Ref. \cite{sapir} and
get $\delta^{\rm recoil}\sim -(3\alpha/\pi) (m_{\mu}/m_p) \ln
(m_{\mu}/m_p) \sim 1.7\ 10^{-3}$. The proton polarizability
correction has been evaluated using the same methods as for
hydrogen: $\delta^{\rm pol}=(4.6\pm0.8)\times10^{-4}$
\cite{cherednikova}. The term $\delta^{\rm
hvp}$ describing the hadron vacuum polarization was
shown to contribute by approximately 20 ppm
\cite{faustov-vp}, unlike
hydrogen where it does not give
any considerable contribution in hydrogen.

With the content of Table \ref{tab1} in mind, we are now ready to
discuss the information that would be provided by measurements of
the hyperfine splitting of the ground state of muonic hydrogen
atoms.

The proton structure correction $\delta^{\rm str}$ in
muonic hydrogen is enhanced (compared to hydrogen) by a factor of $2\ 10^2$.
Therefore, a measurement of $\Delta E^{\rm HFS}$ in
$(\mu^-p)_{1s}$ can not be a good test of QED since QED effects
are overshadowed by the proton structure corrections.
Further on, in both hydrogen and muonic hydrogen, the proton
structure corrections $\delta^{\rm str}$ is dominated by two
independent terms:  the Zemach term $\delta^{\rm rigid}$ and
the polarizability term $\delta^{\rm pol}$. While the Zemach term is
directly related to a well defined physical parameter - the Zemach
radius of the proton $R_p$ (see Eq.~\ref{zemach}), $\delta^{\rm
pol}$ is expressed in terms of the form factors and polarized
structure functions of the proton in an indirect and
case-dependent way and is not associated with a single
parameter.
Compared to hydrogen, both these terms scale approximately as
$(m_{\mu}/m_e)$. This all brings us to the conclusion that
opposite to what was believed by some authors \cite{pla,wein}, the
measurements of $\Delta E^{\rm HFS}$ in hydrogen and muonic
hydrogen atoms {\em are not complementary} in a sense which would
let us extract the values of two universal parameters of the
proton, characterizing its charge and magnetic distribution  and
polarizability.
However,
if assuming that all terms in the right-hand side of Eq.~\ref{zemexp}
are evaluated correctly by theory,
these measurements may be regarded as
repeated experimental determination of the Zemach radius of the
proton.
While the discutible point in this assumption is the credibility ot the
theoretical evaluation of $\delta^{\rm pol}$ (if neglecting
$\delta^{\rm hvp}$),
the repeated measurements of $R_p$ in hydrogen and muonic hydrogen
are the best way to verify it: compatible values  of $R_p$ extracted
from the hyperfine splitting in hydrogen and muonic hydrogen will
confirm the reliability of the theoretical values of $\delta^{\rm pol}$
and vice versa.

The accuracy of $R_p$
depends on the uncertainty of $\delta^{\rm pol}$;
a measurement of the hyperfine splitting of the ground state of
muonic hydrogen based on the results of \cite{cherednikova}
would give the value of $R_p$ accurate to 1\%.
This would be more precise than the value obtained in the
previous section
because of the smaller relative uncertainty of the theoretical
uncertainty of $\delta^{\rm pol}$ (see Table \ref{tab1}), however,
things may change with the more refined
theoretical results to come in the future.
As already mentioned, such an accuracy
would fairly allow to filter the numerous theoretical estimates of $R_p$
and detect a deviation of $G_E/G_M$
from 1 by distinguishing the values of $R_p$
obtained with and without account of
the JLab experimental results \cite{gayou}.
It would be preferable for this purpose
to have the value of $R_p$ accurate to 0.5\% or
better, that requires in turn that the theoretical uncertainty of
$\delta^{\rm pol}$ be brought below $3\,10^{-5}$ and that the experimental
error of $\Delta E^{\rm HFS}_{exp}$ not exceed 30 ppm.

The muonic hydrogen Lamb shift experiment, currently in progress at PSI
\cite{kottmann}, may provide at a later stage
as a by-product the hyperfine
splitting of the $2S$-state of the muonic hydrogen atom with a
relative accuracy of the order of
$0.5\,10^{-3}$. According to
\cite{pachucki-LSmH}, the Zemach correction $\delta^{\rm rigid}$
for the $2S$-state is again $\sim 8\,10^{-3}$. Unless the (yet uncalculated)
polarization correction $\delta^{\rm pol}$ happens to be
anomalously large in this case, this measurement
would therefore provide the value of $R_p$
with an accuracy of 5-10\% \cite{kottmann} which is below the
expected accuracy of the measurements in the 1S state.
The significant improvement of the accuracy of the
proton rms charge radius expected from PSI experiment
will not help imcreasing the accuracy of $R_p$ either
because $R_p$ is independent of the
proton rms charge radius.
In order to determine the Zemach radius of the proton
from the hyperfine splitting of muonic hydrogen atoms in the $1s$ or $2s$
states with
accuracy 1\% or better, the experimental error of $\Delta E^{\rm
HFS}_{\exp}$ should not exceed 50 ppm -- a requirement that is not met
by the PSI Lamb shift experiment \cite{kottmann}.
An alternative experimental method, based
on the response of the muon transfer rate from hydrogen to
oxygen to the population of the $(\mu^-p)_{1S}$ para-state
\cite{werth}, was proposed recently \cite{adamczak}. The method
takes advantage of the recent progress in the development of
tunable lasers in the far infrared range around 6.1 $\mu$m
\cite{kottmann-2000}.
The efficiency of the method has been
demonstrated by means of Monte Carlo simulations. The
experimental error limits have not been
discussed; the main source of experimental uncertainty is
expected to be the Doppler broadening of the transition
lines.

\section{Hyperfine splitting of muonic hydrogen molecular ion}

In conclusion, we would like to briefly outline an alternative
possibility for determining the proton Zemach radius
from a measurement of the hyperfine splitting of
the muonic molecular ion $p\mu^-p$.
The hydrogen muonic molecular ions $(p\mu^-p)$ are formed in the
excited ortho-state with orbital momentum $J=1$ in collisions of
muonic hydrogen atoms with hydrogen molecules \cite{ponomarev}:
\begin{equation}
(\mu^-p)+H_2\rightarrow [(p\mu p)_J p\;ee]^*
\end{equation}
The formation rate $\lambda_{p\mu p}$ is proportional to the
hydrogen target number density $\varphi$: $\lambda_{p\mu
p}=\varphi/\varphi_{0}\times 2.2\,10^6\ {\rm s}^{-1}$
\cite{npa384}. At high densities $\lambda_{p\mu p}$ exceeds the
muon weak decay rate $\lambda_0=0.45\,10^6\ {\rm s}^{-1}$ so that
the muons spend most of their life time bound in muonic molecular
ions.
The spin-orbit and spin-spin interactions of the protons and the
muon split the ortho-state with $J=1$ into 5 hyperfine state
labeled with the quantun numbers $(sF)$ of the total spin
$\mbox{\bf s}=\mbox{\bf s}_{p1}+\mbox{\bf s}_{p2}+\mbox{\bf s}_{\mu}$
and the total angular
momentum $\mbox{\bf F}=\mbox{\bf s}+\mbox{\bf J}$ \cite{npa384,aissing}; the para-state
$J=0$ has no hyperfine structure. The energy separation between the
ortho-levels with $s$=1/2 and $s$=3/2 is $\sim$138 meV, while the
separation within each of these groups is two orders of magnitude
smaller.

A measurement of the splitting between the hyperfine states with
$s$=1/2 and $s$=3/2 might be based on the strong dependence of
the rate of spontaneous ortho-para transitions
$\lambda_{op}=\lambda_{op}(s)$ on $s$:
$\lambda_{op}(1/2)=7.2\,10^4 {\rm s}^{-1}$,
$\lambda_{op}(3/2)=0.8\,10^4 {\rm s}^{-1}$.
Since only the ortho-states with $s$=1/2 are initially
populated at high densities, the
observable
ortho-para transition rate is close to $\lambda_{op}(1/2)$.
The idea of the experimental method would be to use a tunable
laser to stimulate transitions from the $s$=1/2 to the $s$=3/2
hyperfine states, for which $\lambda_{op}(3/2)$ is an order of
magnitude smaller, which would result in a resonance drop of the
observable rate of ortho-para transitions.
Though such an experiment would require the development of
tunable far IR narrow band lasers in the range of 8.8 $\mu$m,
it has some significant advantages compared with measurement in
gaseous hydrogen at room tempertures -- suppressed Doppler
broadening, small muon stopping volume, high energy density
of the laser beam,  high rate of laser-
stimulated ortho-para transitions -- which deserve a careful consideration
in future.
As for the theory, the
accuracy of the currently available results on the hyperfine
structure of $(p\mu^-p)_{J=1}$ is limited to the leading order
Breit and Zemach corrections and to the muon anomalous magnetic
moment \cite{yaf,aissing}; therefore, the theoretical uncertainty
is currently about $10^{-4}$. In order to determine $R_p$ to 1\%
from $p\mu^-p$ hyperfine splitting measurements,
next order QED effects have
to be taken into account together with proton polarizability and
molecular ion finite size effects \cite{melezhik,shima} so that
the theoretical uncertainty be reduced by an order of magnitude to
meet an experimental error below 50 ppm. This is a challenging
task since $(p\mu^-p)$ is a bound system of particles with
comparable masses in which three-body relativistic dynamics may
show up yet in the next-to-leading order.

\section{Conclusions}

By assuming that the numerical results for the various
terms in the theoretical expression for the hyperfine splitting
of the ground state of the hydrogen atom are correct (within the
limits of the claimed accuracy), we have determined the value of
the proton Zemach radius $R_p$. We have also demonstrated that
comparison of the experimental value of $R_p$ with theoretical
calculation is a sensitive test of the quality of the approximation of the
proton form factors at low momentum transfer and of the possible
deviation of the ratio of the charge and magnetic form factors
from 1. To verify the credibility of the theoretical evaluation
of proton polarizability effects, we consider the possibility of
measuring the hyperfine splitting of the ground state of muonic
hydrogen atoms or, in a remote perspective, of muonic hydrogen
molecular ions.

\section{Acknowledgment}
The work on the paper was supported by NATO collaborative linkage
grant PST.CLG.978454. D.B. also recognizes the partial support of
grant Phy-1001 of the Bulgarian National Fund for Scientific
Research. The authors are grateful to A.P.~Martynenko and A.~Le Padellec
for helpful discussions.

\end{document}